\documentclass[12pt,a4paper]{article} 
\usepackage{bbm}
\usepackage[dvips]{graphicx}
\def\ppall{\mathaccent23p}

\long\def\symbolfootnote[#1]#2{\begingroup\def\thefootnote{\fnsymbol{footnote}}\footnote[#1]{#2}\endgroup} 

\begin{document}

\begin{flushright}
 DESY 08-161\\
 HU-EP-08/55\\
 SFB/CPP-08-89
\end{flushright}

\begin{center}
\Large
A comparison of the cut-off effects for\\
Twisted Mass, Overlap and Creutz fermions\\
at tree-level of Perturbation Theory
\symbolfootnote[1]{Presented at the 48. Cracow School of Theoretical Physics: Aspects of Duality, Zakopane, Poland, 13-22 June 2008.}%

\normalsize

\vspace{0.6cm}

KRZYSZTOF CICHY\\
\emph{Adam Mickiewicz University, Faculty of Physics,\\
Umultowska 85, 61-614 Poznan, Poland}\\

\vspace{0.3cm}

J\'ENIFER GONZ\'ALEZ L\'OPEZ\\
\emph{Humboldt--Universit\"at zu Berlin, Institut f\"ur Physik,\\
Newtonstrasse 15, 12489 Berlin, Germany}\\and\\
\emph{DESY Zeuthen, Platanenallee 6, D-15738 Zeuthen, Germany}\\
%\and
%K. Jansen
%\address{DESY Zeuthen, Platanenallee 6, D-15738 Zeuthen, Germany}

\vspace{0.3cm}

AGNIESZKA KUJAWA\\
\emph{Adam Mickiewicz University, Faculty of Physics,\\
Umultowska 85, 61-614 Poznan, Poland}\\
%\and
%A. Shindler
%\address{Theoretical Physics Division, Dept. of Mathematical Sciences,
%\\University of Liverpool, Liverpool L69 7ZL, UK}

%\maketitle
\end{center}

\begin{abstract}
\noindent In this paper we investigate the cutoff effects at tree-level of perturbation theory
for three different lattice regularizations
%discretizations
of fermions
%used in Lattice Field Theory
-- maximally twisted mass Wilson, 
overlap and Creutz fermions. 
We show that all three kinds of fermions exhibit the expected $O(a^2)$ scaling behaviour
in
%with
the lattice spacing. 
Moreover, the size of these cutoff effects
for the considered quantities i.e. the pseudoscalar correlation function $C_{PS}$,
the mass $m_{PS}$ and the decay constant $f_{PS}$
is comparable for all of them.
%\emph{For a fixed value of the quark mass} 
%we also discuss the dependence of the cutoff effects on the quark mass
%and the observable under consideration.
%\emph{for these observables.}
%{\bf I would really remove this last sentence (For a fixed...) completely.}
\end{abstract}
\begin{center}
PACS numbers: 11.15.Ha, 12.38.Gc\\
\end{center}

\newpage
  
\section{Introduction}
The main
%motivation
goal 
of Lattice
Field Theory
%{\bf QCD}
is to study the
non-perturbative aspects of
%{\bf the field theory of strong interactions,}
quantum field theories, in particular
Quantum Chromodynamics (QCD).
%{\bf and, at the moment, the only possible way.}
For example, Lattice QCD is a regularization of QCD
which consists in putting the theory on a four dimensional lattice (discretization)
with lattice spacing $a$, whose inverse is the ultraviolet cutoff of the theory.
%In order to put a theory on a lattice, one has to discretize all relevant degrees of freedom.
The discretization of bosons is relatively straightforward, but when one
tries to discretize fermions in the
%easiest
naive way, the
notorious
%{\bf well-known}
fermion doubling problem emerges -- instead of one fermion
in the continuum limit,
one has as many as $2^d$ fermions, where $d$ is the space-time dimensionality.
%There are several methods to overcome this problem.
As it was originally
%done
proposed
by Wilson \cite{wilson},
the doubling problem can be solved if a different fermionic discretization is chosen,
%which in this case gave place to
the so called Wilson fermions.
After Wilson's proposal, many
%different
alternative fermion
discretizations which remove the doubling problem
%appeared
have been suggested
and are still appearing.
However, as stated by the Nielsen-Ninomiya theorem \cite{nielsen},
new problems will always appear when removing the doublers;
%However, due to the Nielsen-Ninomiya theorem \cite{nielsen},
in order
to eliminate the fermion doubling problem one has to pay the price of either
%explicit
explicitely breaking
%of
chiral symmetry (even in the massless limit), or giving
up locality or translational invariance.
%Given the importance of the latter, the Nielsen-Ninomiya's theorem seems to suggest that one has a choice between chiral symmetry and no fermion doublers.

Much of the effort of lattice QCD goes into finding a lattice theory
without doublers which keeps the
%biggest
largest
possible number of symmetries,
and at the same time reaches the continuum limit as fast as possible
(the dependence on the inverse cutoff, $a$, is as small as possible e.g.
$\mathcal O(a^2)$ leading cutoff dependence is better than $\mathcal O(a)$).

In this paper, we investigate the cutoff effects at tree-level of perturbation theory
of three different discretizations of fermions
-- twisted mass Wilson fermions at maximal twist (MTM), overlap fermions and Creutz fermions,
at a fixed value of the physical quark mass.
We have presented a similar analysis for a different value of the quark mass in \cite{cichy}.
Here we compare the results.
%Thus, here we conclude about the dependence of the cutoff effects on the quark mass.

The MTM fermions \cite{frezzotti01}, \cite{frezzotti04} are relatively cheap to simulate
and they are
by now a
%one of the most
widely used
%approaches to discretizing fermions
fermion discretization.
Although similar to Wilson fermions, they retain a subgroup of chiral 
symmetry which guarantees automatic $\mathcal O(a)$ improvement,
i.e. $\mathcal O(a^2)$ leading cutoff effects.
The price to pay to have a residual chiral symmetry,
is to break a subgroup of the isospin symmetry
transformation\footnote{For a recent review of twisted mass fermions see \cite{shindler}.}.

%Although chiral symmetry is explicitely broken, since they are a kind of Wilson fermions,
%automatic $\mathcal O(a)$ improvement i.e. , is guaranteed.
%They are chirally improved, but still chiral symmetry is explicitly broken.
%Their disadvantage is the explicit breaking of isospin symmetry

%Overlap fermions \cite{neuberger} seem to contradict the Nielsen-Ninomiya theorem
%-- there is no fermion doubling and no chiral symmetry breaking.
%However, as has been shown 
It has been shown by Ginsparg and Wilson \cite{ginsparg} that
there is a way to preserve chiral symmetry on the lattice,
even without the doubler modes, if the corresponding Dirac operator obeys a relation now called the Ginsparg-Wilson relation.
It is a non-standard realization of chiral symmetry \cite{luscher},
because the Dirac operator no longer anticommutes with $\gamma_5$ at non-zero lattice spacing,
but it only anticommutes with a lattice modified version of $\gamma_{5}$.
A particularly simple form of a Dirac operator that obeys the Ginsparg-Wilson relation has been found by Neuberger \cite{neuberger}.
The main disadvantage of overlap fermions is that they are much more costly to simulate
-- by a factor of 30-120 in comparison with MTM fermions\footnote{For a review of overlap fermions see e.g. \cite{niedermayer}.
For a comparison with twisted mass fermions see \cite{goingchiral}.}.

The recently proposed Creutz fermions \cite{creutz} represent the class of minimally-doubled fermions\footnote{Other examples of minimally-doubled fermions were given by Karsten \cite{karsten} and Wilczek \cite{wilczek}.}.
They describe two flavours of quarks and preserve exact chiral symmetry.
However, they break a number of discrete symmetries and isospin symmetry
and this would make their simulation very difficult\footnote{For a discussion of this aspect of Creutz fermions see \cite{bedaque}.}.

\section{Setup}
\subsection{Correlation functions}
In order to
%To
investigate
the cutoff dependence of the
pseudoscalar meson mass and decay constant we have to
first
calculate the correlation function corresponding to this
meson\footnote{For a pedagogical introduction to the methods we have used in this work see \cite{lopez}.}. 
Despite the fact that we work at tree-level of perturbation theory,
we will
refer to the pseudoscalar meson
as the `pion'.

The charged pions are desribed by the following interpolating operator:
\begin{equation}
{\mathcal P}^{\pm}(x)\equiv {\mathcal P}^{1}(x)\mp i {\mathcal P}^{2}(x), 
\label{eq:ppm}
\end{equation}
where the pseudoscalar density ${\mathcal P}^{a}(x)=\bar{\psi}(x)\gamma_{5}\frac{\tau^{a}}{2}\psi(x)$ (for
$a=1,2,3$) and $\tau^a$ are the Pauli matrices.

The time dependence of the correlation function $C_{PS}(t)$ is thus given by:
\begin{equation}
%C(t)=-\sum_{\vec{x}}\langle 0|{\mathcal P}^+(x){\mathcal P}^-(x)|0\rangle.
C_{PS}(t)=-\sum_{\vec{x}}\langle 0|{\mathcal P}^+(x){\mathcal P}^-(0)|0\rangle.
\end{equation} 
%Evaluating this expression,
Performing all the possible Wick's contractions,
one obtains
%a general
the
dependence of the pion correlation function on the quark propagator,
$S_\mu(p)$, given by
\begin{equation}
C_{PS}(t)=\frac{N_c N_d}{L^3 T^2} \sum_{p_4} \sum_{p_4'} \sum_{\vec{p}}
 \sum_{\mu} e^{i(p_4-p_4')t} S_\mu(\vec{p},p_4) S_\mu^*(\vec{p},p_4').
\label{eq:pscorrelator}
\end{equation}
%where
$N_c$ denotes the number of colours and $N_d$ the number of Dirac components.
$L=aN$
is the physical
extent of the lattice in
the spatial directions ($N$ is the number of lattice sites in all spatial directions)
and
$T=aN_4$ the physical extent in the temporal direction ($N_4$ -- the number of lattice points in the time direction).
The possible choices of the index $\mu$ will be explained below (see Eq.~(\ref{eq:qpd})).
%For overlap and Creutz fermions $\mu = U,1,2,3,4$ and for twisted mass fermions $\mu = U,1,2,3,4,5$, and the definition of the propagator components in momentum space $S_\mu(p)$ is given in the next subsection.
The numerical computation of correlation functions consists in directly
evaluating the expression (\ref{eq:pscorrelator}).

\subsection{Quark propagators}

We present in this section the analytical expressions of the quark propagators,
in momentum space and at tree-level of perturbation theory,
for the three kinds of lattice fermions considered in our analysis.\\
%Now, we present the expressions for momentum space propagators at tree-level of perturbation theory for the lattice fermions we employ in numerical computations.

\noindent \textbf{Wilson twisted mass fermions}
\begin{equation}\label{tmprop}
S_{\rm tm}(p)= \frac{-i \ppall_\mu \gamma _{\mu}\mathbbm{1}_{f} +  M(p)
  \mathbbm{1}\mathbbm{1}_{f}-i\mu _{q}\, \gamma _{5}\tau _{3} }
{\sum_\mu \, \ppall_\mu^2 + M(p)^2 + \mu _{q}^{2}},
\end{equation}
where:
\begin{equation}
\ppall_\mu = \frac{1}{a}\sin (ap_{\mu}) , \quad \hat{p}_\mu
=\frac{2}{a}\sin(\frac{ap_{\mu}}{2}) , \quad M(p) =  m_0 + \frac{a}{2}\,
\sum_{\mu}\, \hat{p}_\mu^2 ,
\end{equation}
$\mathbbm{1}$ and $\mathbbm{1}_{f}$ are the identity matrices in Dirac
and flavour space, respectively.
$\mu_q$ is the twisted quark mass and $m_0$ the untwisted quark mass.
The maximal twist setup consists in setting the untwisted mass to
zero\footnote{This can be done exactly only at tree-level. A fine tuning is required in the interacting theory.}
such that the quark mass is only given by the twisted mass.
% -- set to zero to obtain maximal twist at tree-level. 

\vspace{0.2cm}
\noindent \textbf{Overlap fermions}
\begin{equation}
S_{\rm ov}(p)=\frac{-i(1-\frac{ma}{2})F(p)^{-1/2}\ppall_\mu \gamma_\mu + 
\mathcal{M}(p)\mathbbm{1}}{(1-\frac{ma}{2})^2 F(p)^{-1}\sum_\mu\ppall_\mu^2+\mathcal{M}(p)^2}, 
\end{equation}
where:
\begin{equation}
 F(p)=1+\frac{a^4}{2}\sum_{\mu<\nu}\hat{p}_\mu^2 \hat{p}_\nu^2,
\end{equation} 
\begin{equation}
\mathcal{M}(p)=\frac{1}{a}\Bigg(1+\frac{ma}{2}-\Big(1-\frac{ma}{2}\Big)
F(p)^{-1/2}\Big(1-\frac{a^2}{2}\sum_\mu \hat{p}_\mu^2\Big)\Bigg),
\end{equation} 
$m$ is the bare overlap quark mass.

\vspace{0.2cm}
\noindent \textbf{Creutz fermions}

\begin{equation}\label{eq:contCreutz}
S_{\rm C}(p)=
\frac{- i\, \sum_\mu \, p_\mu\, \bar{\gamma}_\mu + m_0\,\mathbf{1}}
{\sum_\mu \sum_\rho \, p_\mu p_\mu\, \bar{a}_{\rho \mu } \bar{a}_{\rho \mu } + 
\sum_{\mu \ne \nu} \sum_\rho \, p_\mu p_\nu\, \bar{a}_{\rho \mu } \bar{a}_{\rho \nu } + m_0^2},
\end{equation}
where:
\begin{displaymath}
\bar{a} = \frac{1}{R}
\left(\begin{array}{cccc}
 1            &  1            & -1            & -1\\
 1            & -1            & -1            &  1\\
 1            & -1            &  1            & -1\\
-\frac{3\sqrt{1-C^2}}{C} & -\frac{3\sqrt{1-C^2}}{C} & -\frac{3\sqrt{1-C^2}}{C} & -\frac{3\sqrt{1-C^2}}{C}
\end{array}\right),
\end{displaymath}
$\bar{\gamma}=\bar{a}^T\gamma$, $m_0$ -- bare quark mass, $C$ -- lattice geometry parameter, $R$ -- $C$-dependent normalization factor needed to obtain the correct continuum limit. We consider two values of $C$: $C=3/\sqrt{10}$ ($R=2$)\footnote{This value corresponds to the hypercubic lattice.} and $C=3/\sqrt{14}$ ($R=2\sqrt{2}$)\footnote{This value corresponds to a highly symmetric lattice geometry, which is the 4-dimensional analogue of graphene structure.}.

We also consider a modification of Creutz's action suggested by Borici \cite{borici}. We call the corresponding fermions the 'Borici fermions' and the quark propagator for them is:
\begin{equation}
\label{prop}
 S_{\rm B}(p)=\frac{-i\sum_\mu G_\mu(ap) \gamma_\mu + m_0\,\mathbbm{1}}{\sum_\mu G_\mu(ap)^2+m_0^2},
\end{equation} 
where the functions $G_\mu(ap)$ are:
\begin{eqnarray}
\label{def-g1}
 G_1(ap)&=&\ppall_1 -\frac{a}{4}\left[ \hat{p}_1^2 + \hat{p}_2^2 - \hat{p}_3^2 - \hat{p}_4^2 \right],\nonumber\\
%\end{equation} 
%\begin{equation} 
\label{def-g2} 
 G_2(ap)&=&\ppall_2 -\frac{a}{4}\left[ -\hat{p}_1^2 + \hat{p}_2^2 - \hat{p}_3^2 - \hat{p}_4^2 \right],\nonumber\\
%\end{equation} 
%\begin{equation} 
\label{def-g3} 
 G_3(ap)&=&\ppall_3 -\frac{a}{4}\left[ -\hat{p}_1^2 - \hat{p}_2^2 + \hat{p}_3^2 - \hat{p}_4^2 \right],\nonumber\\
%\end{equation} 
%\begin{equation} 
\label{def-g4}
 G_4(ap)&=&\ppall_4 - \frac{a}{4}\left[ -\hat{p}_1^2 - \hat{p}_2^2 - \hat{p}_3^2 + \hat{p}_4^2 \right]\nonumber.
\end{eqnarray}

\vspace{0.2cm}
\noindent \textbf{Quark propagator decomposition}.
All of the quark propagators are matrix expressions that can be decomposed in terms of the gamma matrices and the identity matrix:
\begin{equation}
\label{eq:qpd}
S(p)= S_{U}(p)\mathbbm{1} + \sum_{\mu}S_{\mu}(p)\gamma_{\mu}, 
\end{equation}
where $\mu=1,2,3,4$ for overlap and Creutz fermions and $\mu=1,2,3,4,5$ for twisted mass fermions.

\section{Scaling tests}
In this section we present the scaling tests performed on the pseudoscalar correlation functions, masses and decay constants.
We employ the following strategy; we fix $Nm=0.8$ (where $m$ is the bare quark mass in lattice units)
and calculate the correlators
towards the continuum limit ($a\rightarrow 0$).
%for different combinations of $N$ and $m$,
At tree level of perturbation theory this is equivalent to the limit $N\rightarrow\infty$.
%increasing $N$ towards the continuum limit $N\rightarrow\infty$.
The time extent is always set to be larger than and
%proportional to
an integer multiple of
the spatial extent. 

\begin{figure}
\includegraphics[width=0.6\textwidth,angle=270]{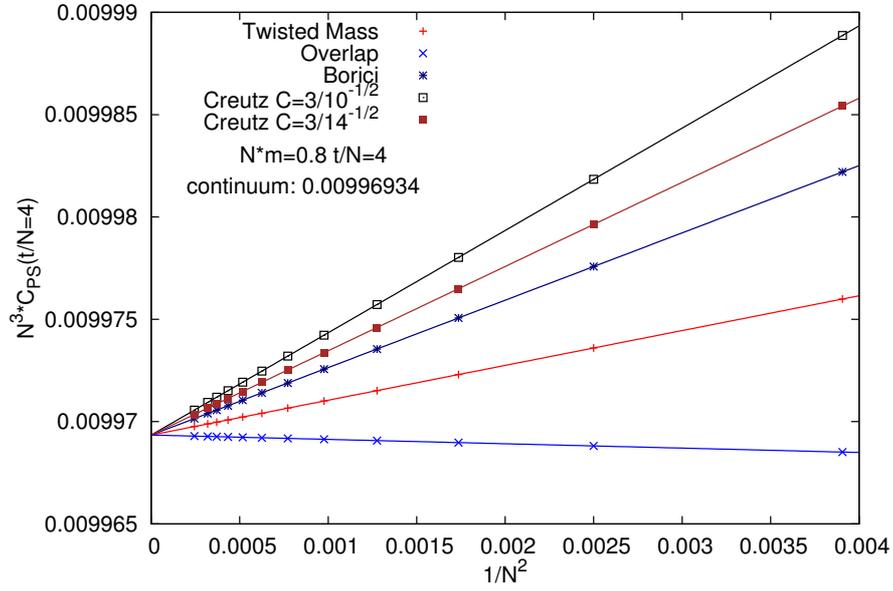}
\caption{Cutoff effects and continuum limit of the pseudoscalar correlation function. \label{fig:corr}}
\end{figure}

In Fig. \ref{fig:corr} we show the correlation function at a fixed physical time $t/N=4$, which
is large enough to allow
%allows
for a reliable extraction of the ground state contribution.
To compare different fermion discretizations, we extract the coefficients, Table \ref{tab:cps},
of the fitting curves shown in Fig. \ref{fig:corr}.
We use the following form of the fitting function:
\begin{equation}
N^3C_{PS} = a + b\frac{1}{N^2} + c\frac{1}{N^4}.
\end{equation}

\begin{figure}
\includegraphics[width=0.6\textwidth,angle=270]{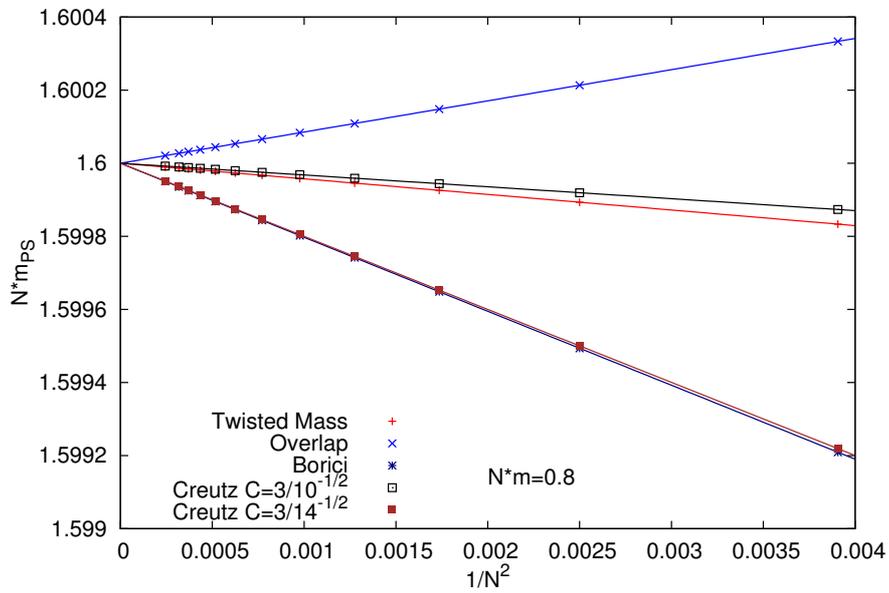}
\caption{Cutoff effects and continuum limit of the pseudoscalar mass. \label{fig:mps}}
\end{figure}

Fig. \ref{fig:mps} shows the pseudoscalar mass and in Table \ref{tab:mps} we have gathered the coefficients of the following fit:
\begin{equation}
Nm_{PS} = a + b\frac{1}{N^2} + c\frac{1}{N^4}.
\end{equation}

\begin{figure}
\includegraphics[width=0.6\textwidth,angle=270]{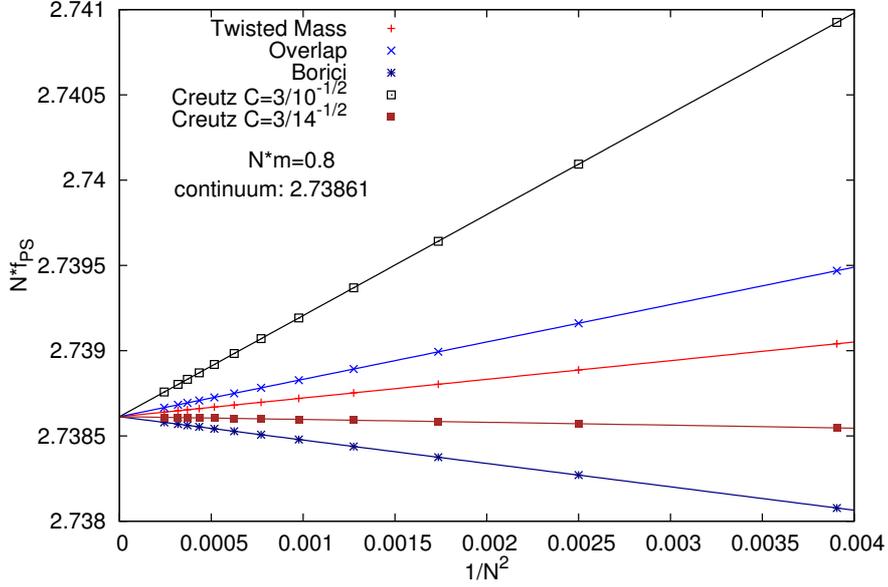}
\caption{Cutoff effects and continuum limit of the pion decay constant. \label{fig:fps}}
\end{figure}

Fig. \ref{fig:fps} presents the pseudoscalar decay constant and Table \ref{tab:fps} the coefficients of the following fit:
\begin{equation}
Nf_{PS} = a + b\frac{1}{N^2} + c\frac{1}{N^4}.
\end{equation}

\begin{table}
\begin{footnotesize}
\begin{center}
\begin{tabular}[c]{|c|| r @{.} l | r @{.} l | r @{.} l |}
\hline
   $N^3C_{PS}(t/N=4)$        & \multicolumn{2}{|c|}{a} &  \multicolumn{2}{|c|}{b} &  \multicolumn{2}{|c|}{c}  \\
\hline
\hline
MTM       & $ 0$&$00996934 $ & $ 0$&$00170143 $ & $ 0$&$00002268 $ \\
\hline
OVERLAP   & $ 0$&$00996934 $ & $-0$&$00021268 $ & $-0$&$00006924 $ \\
\hline
BORICI    & $ 0$&$00996934 $ & $ 0$&$00329653 $ & $-0$&$00116956 $ \\
\hline
CREUTZ -- $C=3/\sqrt{10}$ & $ 0$&$00996934 $ & $ 0$&$00499799 $ & $0$&$000201048 $  \\
\hline
CREUTZ -- $C=3/\sqrt{14}$ & $ 0$&$00996934 $ & $ 0$&$00412066 $ & $ -0$&$00143986 $ \\
\hline
\end{tabular}
\caption{Fit coefficients for the pseudoscalar correlation function.}
\label{tab:cps}
\end{center}
\end{footnotesize}
\end{table}

\begin{table}
\begin{footnotesize}
\begin{center}
\begin{tabular}[c]{|c||c| r @{.} l | r @{.} l |}
\hline
   $Nm_{PS}$       & a  & \multicolumn{2}{|c|}{b}  & \multicolumn{2}{|c|}{c} \\
\hline
\hline
MTM       & $1.6 $ & $-0$&$042667 $ & $ 0$&$003045 $ \\
\hline
OVERLAP   & $1.6 $ & $ 0$&$085333 $ & $ 0$&$008182 $ \\
\hline
BORICI    & $1.6 $ & $-0$&$202667 $ & $ 0$&$058999  $ \\
\hline
CREUTZ -- $C=3/\sqrt{10}$ & $1.6 $ & $-0$&$032000 $  & $-0$&$106063$ \\
\hline
CREUTZ -- $C=3/\sqrt{14}$ & $1.6 $ & $-0$&$200000  $ &  $0$&$029668  $ \\
\hline
\end{tabular}
\caption{Fit coefficients for the pseudoscalar mass.}
\label{tab:mps}
\end{center}
\end{footnotesize}
\end{table}

\begin{table}
\begin{footnotesize}
\begin{center}
\begin{tabular}[c]{|c|| r @{.} l | r @{.} l | r @{.} l |}
\hline
    $Nf_{PS}$       & \multicolumn{2}{|c|}{a} &  \multicolumn{2}{|c|}{b} &  \multicolumn{2}{|c|}{c}   \\
\hline
\hline
MTM       & $ 2$&$73861 $ & $ 0$&$109545 $ & $ -0$&$004307 $  \\
\hline
OVERLAP   & $ 2$&$73861 $ & $ 0$&$219089 $ & $  0$&$028606 $ \\
\hline
BORICI    & $ 2$&$73861$ & $  -0$&$136931 $ & $ -0$&$027123 $  \\
\hline
CREUTZ -- $C=3/\sqrt{10}$ & $ 2$&$73861 $ & $ 0$&$593370   $ & $ -0$&$388244 $  \\
\hline
CREUTZ -- $C=3/\sqrt{14}$ & $ 2$&$73861$ & $ -0$&$015977 $ & $ -0$&$195636  $ \\
\hline
\end{tabular}
\caption{Fit coefficients for the pseudoscalar decay constant.}
\label{tab:fps}
\end{center}
\end{footnotesize}
\end{table}

All types of fermions show the expected behaviour in the lattice spacing -- $\mathcal O(a^2)$ scaling violations.
This is due to the exact chiral symmetry for overlap and Creutz fermions, and to the residual chiral symmetry for MTM.
%The automatic
%$\mathcal O(a)$-improvement in case of twisted mass is only achieved at maximal twist,
%and can be shown to result from symmetry considerations.

The continuum limit for each observable is always the same for every discretization
(and the expected one for the mass at tree-level of perturbation theory),
thus providing a first
check of
%the
consistency of
the corresponding lattice regularizations here analyzed.
%this approach.
The magnitude of the $\mathcal O(a^2)$ effects is, however, very different for different discretizations and depends on the observable under consideration. For example, for the correlation function at a fixed physical time, the smallest effects are exhibited by overlap fermions and the largest by Creutz fermions with $C=3/\sqrt{10}$. For the pion mass, however, the $\mathcal O(a^2)$ scaling violations are the smallest for Creutz fermions with $C=3/\sqrt{10}$ and the largest for Borici fermions.
Therefore, there are no
definite
conclusions, from this scaling test at tree-level,
of which fermions exhibit the smallest $\mathcal O(a^2)$ effects.
%unambiguous conclusions from the scaling tests.
%A conclusion that one discretization leads to uniformly smallest/largest scaling violations can not be reached.
The only clear regularity that we observe is that the disretization errors for twisted mass fermions at maximal twist
are
rather small
%not very large
for all observables that we have considered.
Even the recently proposed Creutz fermions and their modification
by Borici, which break a number of important discrete symmetries,
do not suffer from very large $\mathcal O(a^2)$ scaling violations
at tree-level of perturbation theory and thus can not be excluded from this point of view.
%This is encouraging and
%Only a test in the interacting theory can conclude further.

%{\bf Old version:\\}
%When we compare with our earlier results of Ref. \cite{cichy} (for $Nm=0.5$), we conclude that the fitting coefficients for all the quantities have changed. Therefore, the relative sizes of the cutoff effects do depend on the quark mass.
%When we compare the plots, we can also notice that the curves on the plots of the pseudoscalar mass and decay constant are in the same position and the curves on the plot of the correlation functions at the given timeslice are in different position relatively to each other.\\

%New version:\\
We have compared the results for the scaling behaviour here presented
with the ones discussed in~\cite{cichy},
where the same study for a different value of the quark mass ($Nm=0.5$) was performed.
We observe that, as expected,
for all the quantities which have a well-defined continuum limit,
the relative discretization errors
(ratio of the coefficient of the $\mathcal O(a^2)$ effects with respect to the continuum value)
decrease when decreasing the quark mass independently of the action considered,
while the difference of the relative discretization errors can vary between the actions
considered here when changing the quark mass.

%When we compare with our earlier results of Ref. \cite{cichy} (for $Nm=0.5$), we conclude that the fitting coefficients for the correlation function are not simply related for the cases $Nm=0.5$ and $Nm'=0.8$.
%For example, the smallest $\mathcal O(a^2)$ effects in the former
%occur
%are
%for Borici fermions and in the latter for overlap fermions. This means that the relative size of the $\mathcal O(a^2)$ effects depends also on the quark mass.
%However, the fitting coefficients for the pseudoscalar mass and decay constant are related in a simple way. If we define the ratio of the quark masses $r\equiv Nm'/Nm=1.6$, the coefficients $a'$, $b'$ and $c'$ of the mass and decay constant fits for $Nm'=0.8$ are related to the coefficients for the case $Nm=0.5$ in the following way: for the mass $a'/a=r$, $b'/b=r^3$, $c'/c=r^5$ and for the decay constant $a'/a=r^{-1/2}$, $b'/b=r^{3/2}$, $c'/c=r^{7/2}$. Therefore, we can conclude that the relative size of the cutoff effects for the pseudoscalar mass and decay constant does not depend on the quark mass.\\

\section{Conclusion}
We have performed a scaling test of three different
lattice fermion regularizations
%discretizations
at tree-level of perturbation theory;
%including
the widely used twisted mass and overlap fermions and also the recently proposed minimally-doubled Creutz fermions.
All
these
discretizations lead to the same continuum limit and are $\mathcal O(a)$-improved,
but the relative sizes of
the
$\mathcal O(a^2)$ effects depend strongly on the observable we choose for
the
analysis.
%However, they do not depend on the quark mass.
Therefore, we can not exclude
or put a preference,
from the
%at
tree-level study of the lattice artifacts,
on any particular fermion discretization
of the three here considered.
%and only tests
It will therefore be interesting to test these discretizations
in the interacting theory
%can conclude about the real effects
in practical simulations.

\section*{Acknowledgments}
We want to thank very much K. Jansen and A. Shindler
for their constant support, guidance and many interesting discussions and comments.
And because without them this work would not have been possible.

\end{document}